%
%
%
%
%
%
\magnification=1200
\hsize=6.5truein
\vsize=9.0truein

\baselineskip=1.2\baselineskip
\tolerance=10000
\catcode`\@=11 
 
\def\nolabels{\def\wrlabel##1{}\def\eqlabel##1{}\def\reflabel##1{}}
\def\writelabels{\def\wrlabel##1{\leavevmode\vadjust{\rlap{\smash%
{\line{{\escapechar=` \hfill\rlap{\sevenrm\hskip.03in\string##1}}}}}}}%
\def\eqlabel##1{{\escapechar-1\rlap{\sevenrm\hskip.05in\string##1}}}%
\def\thlabel##1{{\escapechar-1\rlap{\sevenrm\hskip.05in\string##1}}}%
\def\reflabel##1{\noexpand\llap{\noexpand\sevenrm\string\string\string##1}}}
\nolabels
\global\newcount\secno \global\secno=0
\global\newcount\meqno \global\meqno=1
\global\newcount\mthno \global\mthno=1
\global\newcount\mexno \global\mexno=1
\global\newcount\mquno \global\mquno=1
\global\newcount\tblno \global\tblno=1
\def\newsec#1{\global\advance\secno by1 
\global\subsecno=0\xdef\secsym{\the\secno.}\global\meqno=1\global\mthno=1
\global\mexno=1\global\mquno=1\global\figno=1\global\tblno=1

\bigbreak\medskip\noindent{\bf\the\secno. #1}\writetoca{{\secsym} {#1}}
\par\nobreak\medskip\nobreak}
\xdef\secsym{}
\global\newcount\subsecno \global\subsecno=0
\def\subsec#1{\global\advance\subsecno by1 \global\subsubsecno=0
\xdef\subsecsym{\the\subsecno.}
\bigbreak\noindent{\bf\secsym\the\subsecno. #1}\writetoca{\string\quad
{\secsym\the\subsecno.} {#1}}\par\nobreak\medskip\nobreak}
\xdef\subsecsym{}
\global\newcount\subsubsecno \global\subsubsecno=0
\def\subsubsec#1{\global\advance\subsubsecno by1
\bigbreak\noindent{\it\secsym\the\subsecno.\the\subsubsecno.
                                   #1}\writetoca{\string\quad
{\the\secno.\the\subsecno.\the\subsubsecno.} {#1}}\par\nobreak\medskip\nobreak}
\global\newcount\appsubsecno \global\appsubsecno=0
\def\appsubsec#1{\global\advance\appsubsecno by1 \global\subsubsecno=0
\xdef\appsubsecsym{\the\appsubsecno.}
\bigbreak\noindent{\it\secsym\the\appsubsecno. #1}\writetoca{\string\quad
{\secsym\the\appsubsecno.} {#1}}\par\nobreak\medskip\nobreak}
\xdef\appsubsecsym{}
\def\appendix#1#2{\global\meqno=1\global\mthno=1\global\mexno=1
\global\figno=1\global\tblno=1
\global\subsecno=0\global\subsubsecno=0
\global\appsubsecno=0
\xdef\appname{#1}
\xdef\secsym{\hbox{#1.}}
\bigbreak\bigskip\noindent{\bf Appendix #1. #2}
\writetoca{Appendix {#1.} {#2}}\par\nobreak\medskip\nobreak}
%
%
\def\eqnn#1{\xdef #1{(\secsym\the\meqno)}\writedef{#1\leftbracket#1}%
\global\advance\meqno by1\wrlabel#1}
\def\eqna#1{\xdef #1##1{\hbox{$(\secsym\the\meqno##1)$}}
\writedef{#1\numbersign1\leftbracket#1{\numbersign1}}%
\global\advance\meqno by1\wrlabel{#1$\{\}$}}
\def\eqn#1#2{\xdef #1{(\secsym\the\meqno)}\writedef{#1\leftbracket#1}%
\global\advance\meqno by1$$#2\eqno#1\eqlabel#1$$}
%
%
\def\thm#1{\xdef #1{\secsym\the\mthno}\writedef{#1\leftbracket#1}%
\global\advance\mthno by1\wrlabel#1}
\def\exm#1{\xdef #1{\secsym\the\mexno}\writedef{#1\leftbracket#1}%
\global\advance\mexno by1\wrlabel#1}
%
%
\def\tbl#1{\xdef #1{\secsym\the\tblno}\writedef{#1\leftbracket#1}%
\global\advance\tblno by1\wrlabel#1}
%
\newskip\footskip\footskip14pt plus 1pt minus 1pt 
\def\f@@t{\baselineskip\footskip\bgroup\aftergroup\@foot\let\next}
\setbox\strutbox=\hbox{\vrule height9.5pt depth4.5pt width0pt}
\global\newcount\ftno \global\ftno=0
\def\foot{\global\advance\ftno by1\footnote{$^{\the\ftno}$}}
%
\newwrite\ftfile
\def\footend{\def\foot{\global\advance\ftno by1\chardef\wfile=\ftfile
$^{\the\ftno}$\ifnum\ftno=1\immediate\openout\ftfile=foots.tmp\fi%
\immediate\write\ftfile{\noexpand\smallskip%
\noexpand\item{f\the\ftno:\ }\pctsign}\findarg}%
\def\footatend{\vfill\eject\immediate\closeout\ftfile{\parindent=20pt
\centerline{\bf Footnotes}\nobreak\bigskip\input foots.tmp }}}
\def\footatend{}
%
%
\global\newcount\refno \global\refno=1
\newwrite\rfile
\def\ref{\the\refno\nref}

\def\nref#1{\xdef#1{\the\refno}\writedef{#1\leftbracket#1}%
\ifnum\refno=1\immediate\openout\rfile=refs.tmp\fi
\global\advance\refno by1\chardef\wfile=\rfile\immediate
\write\rfile{\noexpand\item{[#1]\ }\reflabel{#1\hskip.31in}\pctsign}\findarg}
\def\findarg#1#{\begingroup\obeylines\newlinechar=`\^^M\pass@rg}
{\obeylines\gdef\pass@rg#1{\writ@line\relax #1^^M\hbox{}^^M}%
\gdef\writ@line#1^^M{\expandafter\toks0\expandafter{\striprel@x #1}%
\edef\next{\the\toks0}\ifx\next\em@rk\let\next=\endgroup\else\ifx\next\empty%
\else\immediate\write\wfile{\the\toks0}\fi\let\next=\writ@line\fi\next\relax}}
\def\striprel@x#1{} \def\em@rk{\hbox{}}
\def\lref{\begingroup\obeylines\lr@f}
\def\lr@f#1#2{\gdef#1{\ref#1{#2}}\endgroup\unskip}

\def\addref#1{\immediate\write\rfile{\noexpand\item{}#1}} 
\def\startrefs#1{\immediate\openout\rfile=refs.tmp\refno=#1}
\def\xref{\expandafter\xr@f}\def\xr@f[#1]{#1}
\def\refs#1{[\r@fs #1{\hbox{}}]}
\def\r@fs#1{\edef\next{#1}\ifx\next\em@rk\def\next{}\else
\ifx\next#1\xref #1\else#1\fi\let\next=\r@fs\fi\next}
%

%
 \newwrite\ffile\global\newcount\figno \global\figno=1
%
%
\def\fig{\the\figno\nfig}
\def\nfig#1{\xdef#1{\secsym\the\figno}%
\writedef{#1\leftbracket \noexpand~\the\figno}%
\ifnum\figno=1\immediate\openout\ffile=figs.tmp\fi\chardef\wfile=\ffile%
\immediate\write\ffile{\noexpand\medskip\noexpand\item{Figure\ \the\figno. }
\reflabel{#1\hskip.55in}\pctsign}\global\advance\figno by1\findarg}
\def\xfig{\expandafter\xf@g}\def\xf@g \penalty\@M\ {}
\def\figs#1{figs.~\f@gs #1{\hbox{}}}
\def\f@gs#1{\edef\next{#1}\ifx\next\em@rk\def\next{}\else
\ifx\next#1\xfig #1\else#1\fi\let\next=\f@gs\fi\next}
%
%
\newwrite\lfile

{\escapechar-1\xdef\pctsign{\string\%}\xdef\leftbracket{\string\{}
\xdef\rightbracket{\string\}}\xdef\numbersign{\string\#}}

\def\writestop{\def\writestoppt{\immediate\write\lfile{\string\pageno%
\the\pageno\string\startrefs\leftbracket\the\refno\rightbracket%
\string\def\string\secsym\leftbracket\secsym\rightbracket%
\string\secno\the\secno\string\meqno\the\meqno}\immediate\closeout\lfile}}
\def\writestoppt{}\def\writedef#1{}

\def\seclab#1{\xdef #1{\the\secno}\writedef{#1\leftbracket#1}\wrlabel{#1=#1}}

\def\subseclab#1{\xdef #1{\secsym\the\subsecno}%
\writedef{#1\leftbracket#1}\wrlabel{#1=#1}}
\def\appsubseclab#1{\xdef #1{\secsym\the\appsubsecno}%
\writedef{#1\leftbracket#1}\wrlabel{#1=#1}}
\def\subsubseclab#1{\xdef #1{\secsym\the\subsecno.\the\subsubsecno}%
\writedef{#1\leftbracket#1}\wrlabel{#1=#1}}
\newwrite\tfile \def\writetoca#1{}
\def\leaderfill{\leaders\hbox to 1em{\hss.\hss}\hfill}
\def\writetoc{\immediate\openout\tfile=toc.tmp
   \def\writetoca##1{{\edef\next{\write\tfile{\noindent ##1
   \string\leaderfill {\noexpand\number\pageno} \par}}\next}}}

%
\catcode`\@=12 
%
%
%
%
%
\def\dbend{{\manual\char127}}
\def\d@nger{\medbreak\begingroup\clubpenalty=10000
    \def\par{\endgraf\endgroup\medbreak} \noindent\hang\hangafter=-2
    \hbox to0pt{\hskip-\hangindent\dbend\hfill}\ninepoint}
\outer\def\danger{\d@nger}

\def\p{\partial}

\def\darr#1{\raise1.5ex\hbox{$\leftrightarrow$}\mkern-16.5mu #1}
\def\half{{\textstyle{1\over2}}} 

%
%
\def\al{\alpha}
\def\be{\beta}
\def\ga{\gamma}  
\def\de{\delta}

\def\th{\theta}

\def\la{\lambda} 
\def\rh{\rho}
\def\si{\sigma}

\def\om{\omega}  
%
%

%

%
%
\def\cA{{\cal A}}

\def\cL{{\cal L}}

\def\cR{{\cal R}}

%

%

%
%
\def\amsyes{y }

\def\answ{y }

\ifx\answ\amsyes
\input amssym.def


\def\CC{{\Bbb C}}
\def\ZZ{{\Bbb Z}}

\else
\def\ZZ{{Z\!\!\!Z}}              
\def\CC{{I\!\!\!\!C}}

\def\cA{{\cal A}} 
\def\cR{{\cal R}} 
\fi
%

%
%

%
%
%

\def\CMP#1{Comm.\ Math.\ Phys.\ {\bf #1}}

\def\MPL#1{Mod.\ Phys.\ Lett.\ {\bf #1}}

\def\PLB#1{Phys.\ Lett.\ {\bf {#1}B}}

\def\PTP#1{Prog.\ Theor.\ Phys.\ Suppl.\ {\bf #1}}
\def\SMD#1{Sov.\ Math.\ Dokl.\ {\bf {#1}}}



\def\wR{\widehat{R}}
\def\wcR{\widehat{\cR}}  \def\wcA{{\widehat{\cA}}}
\def\pp{{+\!\!\!+}}
\def\iden{{1}}
\def\mywedge{{\textstyle{\bigwedge}}}

%
%

\lref\Man{
Yu.~I.~Manin, {\it Quantum Groups and Non-Commutative Geometry},
Publications CRM, Montr\'eal, 1989; \CMP{123} (1989) 163.}

\lref\SV{
J.H.~Schwarz and P.~van~Nieuwenhuizen, Lett.\ Nuovo Cimento {\bf 34}
(1982) 21.}

\lref\WZ{
J.~Wess and B.~Zumino, Nucl.\ Phys.\ (Proc.\ Suppl.) {\bf 18B} (1990) 302. }

\lref\Zu{
B.~Zumino, in Proc.\ of the XIX International Conference on Group 
Theoretic Methods in Physics, Salamanca, Spain, 1992, 
({\tt hep-th/9212093}).}

\lref\CSZ{
C.~Chryssomalakos, P.~Schupp and B.~Zumino, 
Alg.\ Anal.\ {\bf 6} (1994) 252, ({\tt hep-th/9401141}).}

\lref\Gu{
D.I.~Gurevich, \SMD{33} (1986) 758.}

\lref\Bern{
D.~Bernard, \PTP{102} (1990) 49; \PLB{260} (1991) 389.}

\lref\Zua{
B.~Zumino, \MPL{A6} (1991) 1225.}

\lref\Zub{
W.B.~Schmidte, S.P.~Vokos and B.~Zumino, Zeit.\ f\"ur 
Physik {\bf C8} (1990) 249.}

%
%
%
\hfuzz=15pt
\nopagenumbers
%
\pageno=0
%
%
%
\line{}
\vskip1cm
\centerline{\bf FUSING THE COORDINATES OF QUANTUM SUPERSPACE}
\vskip1cm
 
\centerline{Peter BOUWKNEGT$\,^{1}$, 
Jim McCARTHY$\,^1$ and 
Peter van NIEUWENHUIZEN$\,^2$}
\bigskip
 
\centerline{\sl $^1$ Department of Physics and Mathematical Physics \&}
\centerline{\sl Institute for Theoretical Physics}
\centerline{\sl University of Adelaide}
\centerline{\sl Adelaide, SA~5005, Australia}
\bigskip
 
\centerline{\sl $^2$ Institute for Theoretical Physics }
\centerline{\sl State University of New York}
\centerline{\sl Stony Brook, NY~11794-3800, USA}
\medskip
\vskip1.5cm
 
\centerline{\bf ABSTRACT}\medskip
{\rightskip=1cm 
\leftskip=1cm 
\noindent
We introduce the notion of a fused quantum superplane by allowing 
for terms $\th\th\sim x$ in the defining relations.  We develop 
the differential calculus for a large class of fused quantum 
superplanes related to particular solutions of the Yang-Baxter 
equation.
}
 
\vfil
\line{ADP-96-35/M46 \hfil}
\line{ITP-SB-96-60\hfil}
\line{{{\tt hep-th/9611067}}\hfil November 1996}
 
\eject
\footline{\hss \tenrm -- \folio\ -- \hss}
\baselineskip=1.2\baselineskip

In the conventional approach to quantum superplanes (see e.g.\ [\Man]
for a review) one starts with a set of quadratic relations between 
a set of $M$ bosonic coordinates $x^\mu$ and $N$ fermionic 
coordinates\foot{The distinction between bosonic and fermionic coordinates
may seem artificial in the general context of [\Man], we make it since
we will be imposing the corresponding $\ZZ_2$ grading for physical
reasons.} $\th^\al$ 
\eqn\eqBMVa{ \eqalign{
\th^\al \th^\be ~=~ \wR^{\al\be}_{\ga\de}\th^\ga \th^\de \,,\cr 
x^\mu \th^\al ~=~ \wR^{\mu \al}_{\be \nu} \th^\be x^\nu\,,\cr
x^\mu x^\nu ~=~ \wR^{\mu\nu}_{\rh\si} x^\rh x^\si \,.\cr}
}
On introducing coordinates $z^a=\{x^\mu,\th^\al\}$ 
$(\mu=1,\ldots,M$, $\al=1,\ldots,N$) these may be written as
\eqn\eqBMVb{
z^a z^b ~=~ \wR^{ab}_{cd} z^c z^d \,.
}
The quantum superplane $\cA_q$, determined by $\wR$, is then defined to be 
the quadratic algebra obtained by modding out the relations \eqBMVb\
from the free associative algebra generated by the $z^a$.  One may
consider $\cA_q$ which are a 
flat deformation of the usual $(M|N)$-superspace: 
the matrix $\wR$ then depends on a set of deformation parameters, 
here collectively
denoted by $q$, such that $q\to1$ corresponds to the classical limit
$\wR^{ab}_{cd}(q=1) = (-1)^{|a||b|} \de^{a}_d \de^b_c$; moreover, 
the flatness condition is the statement that $\cA_q$ is isomorphic
to $\cA \otimes \CC(q)$ as a $\CC(q)$-module, where $\cA=\cA_{q=1}$
is the underformed superplane.
In most examples $\wR$ is
taken to be a solution of the Yang-Baxter equation (YBE).

Some time ago, before the advent of quantum groups,
it was suggested that the coordinates
$x^\mu$ and $\th^\al$ of superspace are not on equal footing,
but rather that the coordinates $x^\mu$ of spacetime are ``composites''
of the more fundamental fermionic variables $\th^\al$ [\SV].
To explore this idea a point particle model was 
constructed, with Lagrangian $\cL = \bar\th (\ga\cdot x)^{-1} \dot{\th} + 
\cL_1(x,\dot{x},\th)$, where $\cL_1$ contains terms quadratic in
$\dot{x}$ such that the only second class constraints were those for the 
conjugate momenta of $\th^\al$.
The Dirac brackets for the fermionic coordinates%
\foot{The remaining brackets were $[x^\mu,x^\nu]_D = [x^\mu,\th^\al]_D = 0$,
$[x^\mu,p_\nu]_D=\de^\mu_\nu$, $[\th,p_\mu]_D = {1\over2} 
\ga_\mu (\ga\cdot x)^{-1} \th$ and $[p_\mu,p_\nu]_D = 
{1\over2} \bar\th \ga_{\mu\nu\rh} \th {x^\rh\over x^4}$.}
were found to realize the compositeness of the $x^\mu$
\eqn\eqBMVc{
[ \th^\al, \th^\be ]_D = \ga_\mu^{\al\be} x^\mu \,.
}
However, whereas Lorentz generators were constructed, 
yielding the conventional Lorentz algebra, the authors of [\SV] were
unable to construct translation generators $P_\mu$ satisfying 
$[P_\mu,P_\nu]_D=0$ and acting on the $x^\mu$ as usual,
$[P_\mu,x^\nu]_D=\de_\mu^\nu$.  The proposal was incomplete and 
very speculative, and no further work was done on the subject.

In this letter, we reconsider \eqBMVc\ in the context of quantum 
deformations of the superplane.  Motivated by [\SV] we
generalize \eqBMVb\ by allowing for a term linear in coordinates
\eqn\eqBMVf{
z^a z^b ~=~ \wR^{ab}_{cd}\, z^c z^d + T^{ab}_c z^c \,.
}
Such a term preserves the dimensions $-1$ of $x^\mu$ and 
$-\half$ of $\th^\al$ provided
\eqn\eqBMVe{
T^{ab}_c ~=~ 0\,,\qquad {\rm unless}\ \ a=\al\,, b=\be\,, c=\mu\,,
}
which we will henceforth assume. 
An equation similar to \eqBMVf\ was considered in the context
of braided Lie algebras [\Gu,\Bern], the $z^a$ then corresponding
to the generators of a deformed Lie algebra.  However, in that 
context the $T^{ab}_c$ tend to the structure constants of the Lie
algebra in the classical limit whereas in our approach we allow the
limit to be zero.  Further, the conditions that arise on $\wR$ and $T$ are 
different in the two contexts.
The fused quantum superplane $\cA_q$, determined by $\wR^{ab}_{cd}$ and 
$T^{ab}_c$, is then defined to be 
the quadratic algebra obtained by modding out the relations \eqBMVf\
from the free associative algebra generated by the $z^a$.  Again we
are interested in the case where $\cA_q$ is a flat deformation
of the usual quantum superplane.  
We find that the flatness conditions 
in the fused case are more restrictive than in the unfused case since
by exploring the associativity conditions one is led to 
compatibility relations between the various components of $\wR$.
In addition we impose the natural condition
\eqn\eqBMVd{
\wR^{aa}_{cd} ~=~ (-1)^{|a|} \de^a_c \de^a_d\,.
}

For definiteness we discuss a simple model first.  Consider a $(2|2)$
superspace in $d=(1,1)$ spacetime with lightcone
coordinates $x^\pp, x^=, \th^+$ and $\th^-$.  Assuming the preservation
of the Lorentz index structure (which is a natural assumption if we 
ultimately want to have a deformation of the Lorentz group acting on
our fused quantum superplane) leads to an ansatz with 
eight free parameters.  On  
demanding compatibility of the associativity relations 
such as $((\th^+ \th^+) \th^+) = (\th^+ (\th^+ \th^+))$
with the defining relations \eqBMVf, it is clear that
the complete set may be computed once the fundamental relations
for two $\th$'s have been given.  We find (cf.\ [\Zub])
\eqn\eqBMVz{  \eqalign{
\th^+ \th^- ~=~ q \th^- \th^+\,, \cr
\th^+ \th^+ ~=~ \al x^\pp\,, &\qquad\qquad   \th^- \th^- ~=~ \be x^=\,, \cr
x^\pp \th^+ ~=~  \th^+ x^\pp\,, &\qquad\qquad   
x^= \th^+ ~=~ q^{-2} \th^+ x^= \,,\cr
x^\pp \th^- ~=~ q^2 \th^- x^\pp \,,&\qquad\qquad   
x^= \th^- =  \th^- x^= \,,\cr
x^\pp x^= ~=~ q^4 x^= x^\pp\,.\cr}
}
For $\al\neq0$ or $\be\neq0$, one can scale $x^\pp$ or $x^=$, respectively,
to put the equations in a form with $\al=1$ or $\be=1$.  From this we 
see that the equations only contain one true deformation parameter, $q$.
We will keep the $\al,\be$ explicit, however, in order to discuss the 
ordinary superplane as the limit $\al,\be\to 0$.  Note that the equations
\eqBMVz\ may be written in the form \eqBMVf--\eqBMVd,
where we may fix $T^{++}_{\pp} = 2 \al$ and $T^{--}_{=} = 2 \be$. 
A generalization of the solution \eqBMVz\ to $(N|N)$ superspace is 
easily constructed along the same lines.  We will return to this 
generalization later.

We now return to a more general discussion of the fused quantum superplanes.
First, we may also write \eqBMVf\ as 
\eqn\eqBMVg{
(\de^a_c\de^b_d - \wR^{ab}_{cd}) z^c z^d ~=~ T^{ab}_c z^c \,.
}
In this form it is obvious that the relations \eqBMVf\ are the same
as those corresponding to $(\wR',T')$ given by
\eqn\eqBMVh{ \eqalign{
(\de^a_c\de^b_d - \wR'^{ab}_{cd}) & ~=~ \al^{ab} \,
  (\de^a_c\de^b_d - \wR^{ab}_{cd}) \,,\cr
T'{}^{ab}_c & ~=~ \al^{ab}\, T^{ab}_c \,,\cr}
}
for any choice of $\al^{ab}\neq0$.  A more general equivalence of this
form is obtained by replacing $\al^{ab}$ by an invertible matrix
$\al^{ab}_{a'b'}$.  There are further equivalences corresponding to
linear redefinitions of the coordinates.  Since it is not our aim in
this paper to give the most general quantum superplane up to these
equivalences we will not explore this any further.

For invertible $\wR$-matrices, the relation \eqBMVf\ can also be written
as
\eqn\eqBMVi{
z^a z^b ~=~ (\wR^{-1})^{ab}_{cd} z^c z^d - (\wR^{-1})^{ab}_{cd} 
  T^{cd}_e z^e \,.
}
Although more general fused quantum superplanes are feasible, it is 
natural to impose that the equations \eqBMVi\ and \eqBMVf\ are related by
\eqn\eqBMVj{ \eqalign{
(\wR^{-1})^{ab}_{cd} & ~=~ \la\, \wR^{ab}_{cd} + (1-\la)\, \de^a_c\de^b_d\,,\cr
(\wR^{-1})^{ab}_{cd} T^{cd}_e & ~=~ -\la\, T^{ab}_e \,,\cr}
}
for some $\la\neq0$.  Equivalently, 
\eqn\eqBMVaf{ \eqalign{
\la \wR^2 + (1-\la) \wR - \iden & ~=~ (\la \wR + \iden ) 
  (\wR -\iden ) ~=~ 0\,,\cr
\la \wR T  & ~=~ -T \,,\cr}
}
i.e.\ $\wR$ satisfies a quadratic characteristic equation 
and $T^{ab}_c$ is an eigenvector of $\wR$ with eigenvalue $-1/\la$ 
for all indices $c$.
Moreover, any $(\wR,T)$-system satisfying \eqBMVaf\ is equivalent under 
\eqBMVh\ to an $(\wR,T)$-system satisfying
\eqn\eqBMVk{ 
\wR^2  ~=~ \iden\,, 
}
\eqn\eqBMVl{
\wR T  ~=~ -T \,.
}
Henceforth we will work with the `gauge' \eqBMVk, \eqBMVl.  Together
with \eqBMVd\ we will refer to these conditions as the naturalness
conditions on $(\wR,T)$.  With the above naturalness conditions our
differential calculus takes a particularly simple form, but the
corresponding formulas for the more general gauge \eqBMVaf\ can be easily 
worked out.\foot{The formulas in [\WZ], for the bosonic $2$-plane, correspond
to the gauge $\la = q^2$.}  Again we stress that there are additional 
relations on $\wR$ and $T$ coming from the compatibility of 
associativity of the coordinate ring 
with the defining relations \eqBMVf.  We do not know
how to write these relations in closed formulas.

Both to construct interesting examples and to construct a 
differential calculus on fused quantum superplanes along the lines 
of [\WZ,\Zua,\Zu,\CSZ] it proves convenient to introduce 
an additional bosonic coordinate $z^0$ -- commuting with all the 
other coordinates $z^a$ and which can be consistently 
specialized to a constant (as will be done at a later
stage) -- and rewrite \eqBMVf\ 
in terms of $z^A \equiv \{ z^0, z^a\} = \{ z^0, x^\mu, \th^\al\}$
(cf.\ [\Gu,\Bern])
\eqn\eqBMVaa{
z^A z^B ~=~ \wcR^{AB}_{CD}\, z^C z^D \,,
}
with
\eqn\eqBMVab{
\wcR^{AB}_{CD} ~=~ \left( \matrix{ 
   1 & 0 & 0 & 0 \cr
   0 & 0 & \de^b_c & 0 \cr
   0 & \de^a_d & 0 & 0 \cr
   0 & {1\over2} T^{ab}_d & {1\over2} T^{ab}_c & \wR^{ab}_{cd} \cr} \right)\,.
}
Here the matrix rows correspond to $(00)$, $(0b)$, $(a0)$ and $(ab)$,
the columns similarly to $(00)$, $(0d)$, $(c0)$ and $(cd)$,
respectively.  The quantum superplane $\cA_q$ extended by $z^0$ will
be denoted by $\wcA_q$.
The equations \eqBMVk\ and \eqBMVl\ for $(\wR,T)$ imply that $\wcR$ 
satisfies the same characteristic equation as $\wR$, i.e.\
\eqn\eqBMVn{
\wcR^2 ~=~ \iden \,.
}

A particularly interesting class of fused quantum superplanes is
those for which $\wcR$ satisfies the
Yang-Baxter equation
\eqn\eqBMVac{
(\wcR \otimes \iden ) (\iden \otimes \wcR) (\wcR \otimes \iden ) ~=~
   (\iden \otimes \wcR) (\wcR \otimes \iden ) (\iden \otimes \wcR) \,,
}
or, in components, 
\eqn\eqBMVad{
\wcR^{AB}_{PQ}\, \wcR^{PR}_{KL}\, \wcR^{QC}_{RM} ~=~
  \wcR^{BC}_{PQ} \, \wcR^{AP}_{KR}\, \wcR^{RQ}_{LM} \,.
}
Equation \eqBMVac, together with \eqBMVn, implies that $\wcR$ constitutes
a representation of the permutation group.
In terms of $\wR^{ab}_{cd}$ and $T^{ab}_c$, equation \eqBMVad\ reads
\eqn\eqBMVae{ \eqalign{
\wR^{ab}_{pq}\, \wR^{qc}_{rm}\, \wR^{pr}_{kl} &  ~=~
  \wR^{bc}_{pq} \, \wR^{ap}_{kr}\, \wR^{rq}_{lm} \cr
\wR^{ab}_{lq} T^{qc}_m + \wR^{ab}_{pq} \wR^{qc}_{rm} T^{pr}_l & ~=~ 
  \wR^{aq}_{lm} T^{bc}_q + \wR^{bc}_{pq} \wR^{rq}_{lm} T^{ap}_r \cr
T^{ab}_p T^{pc}_l + \wR^{ab}_{pq} T^{pr}_l T^{qc}_r & ~=~ 
  T^{bc}_q T^{aq}_l + \wR^{bc}_{pq} T^{ap}_r T^{rq}_l \cr
T^{ab}_p T^{pc}_l & ~=~ 2 T^{bc}_p T^{ap}_l + \wR^{bc}_{pq} T^{ap}_r T^{rq}_l 
\cr}
}
The last two of these equations are trivially satisfied because of
the preservation of dimension condition \eqBMVe.
Also note that the YBE is not preserved under the equivalence \eqBMVh.

 It is straightforward to check that the model in \eqBMVz\
(written in the form of equations \eqBMVf--\eqBMVd) satisfies
these YBEs.  In fact, suppose one takes \eqBMVf\ in its most general
form for the $(2|2)$ plane which preserves grading, dimension, and
Lorentz index structure (as discussed previously).  Then imposing the
YBE leaves \eqBMVz\ as the most general solution if one excludes those
solutions (nonflat deformations) which impose spurious vanishing of
products of the generators.  One can push this slightly further.  If
we replace Lorentz index preservation by the naturalness conditions,
and still only allow a deformation with $T^{++}_{\pp}$ and
$T^{--}_{=}$ nonzero, then the YBE still leads to the same unique
solution.  It is not clear how to make this analysis in a more general
case, but this does show a rather pleasing consistency of the idea of
naturalness at least for this case.

Let us consider the case of the $(M|N)$ superplane; i.e.\ a 
plane with coordinates $(x^\mu,\th^\al)$, $\mu=1,\ldots,M$, 
$\al=1,\ldots,N$.
We can look for solutions where the braiding matrix $\wR$ just has the 
effect of exchanging the coordinates up to some multiplicative factor, i.e.\
\eqn\eqBMVcb{
\wR^{ab}_{cd} ~=~ q_{ab} \de^a_d \de^b_c\,,
}
for some set of parameters $q_{ab}$.  The condition \eqBMVk\ leads to 
\eqn\eqBMVcc{
q_{ab} ~=~ q_{ba}^{-1}\,,
}
thus, in particular, with \eqBMVd, $q_{aa} = (-1)^{|a|}$, while 
\eqBMVl\ leads to
\eqn\eqBMVcd{
q_{ab} T^{ba}_c ~=~ - T^{ab}_c\,.
}
The first equation in \eqBMVae\
is now automatically satisfied, while the second equation leads to the 
condition
\eqn\eqBMVce{
q_{ad} ~=~ q_{ab} q_{ac} \,,\qquad {\rm whenever}\ \ T^{bc}_d \neq 0\,.
}
One obvious solution to \eqBMVcc--\eqBMVce\ is the trivial one, i.e.\
\eqn\eqBMVcf{
q_{ab} ~=~ (-1)^{|a| |b|}\,,
}
in which case the only 
constraint on $T^{\al\be}_\mu$ (arising from \eqBMVcd) is 
$T^{\al\be}_\mu=T^{\be\al}_\mu $.  This corresponds 
to the situation studied in [\SV], i.e.\
\eqn\eqBMVcg{ \eqalign{
 \th^\al \th^\be & ~=~ - \th^\be \th^\al + 2 T^{\al\be}_\mu x^\mu \,,\cr
x^\mu \th^\al & ~=~ \th^\al x^\mu \,,\cr
x^\mu x^\nu & ~=~ x^\nu x^\mu \,.\cr}
}

A less nontrivial solution, generalizing \eqBMVz, is to consider an equal 
number of fermionic and bosonic coordinates $(x^i,\th^i)$, 
$i=1,\ldots,N$ and making the assumption that the relations \eqBMVf\ preserve
the dimension in each direction $i$, i.e.\ $T^{\th^i \th^j}_{x^k} =
\al_i \de^i_k \de^j_k$, ($\al_i\neq0$),
while all other $T^{ab}_c$ vanish.  We immediately
arrive at the following solution of \eqBMVcc--\eqBMVce\ 
\eqn\eqBMVca{ \eqalign{
\th^i \th^j & ~=~ - q_{ij} \th^j \th^i + 2 \al_i \de^{ij} x^i \,,\cr
x^i \th^j & ~=~ q_{ij}^2 \th^j x^i\,,\cr
x^i x^j & ~=~ q_{ij}^4 x^jx^i\,,\cr}
}
where $q_{ii}=1$,
$q_{ji} = q_{ij}^{-1}$ and the $q_{ij}$, $i<j$ and $\al_i$ are 
arbitrary deformation parameters.  We will refer 
to this solution as the fused quantum superplane $\cA_{(N|N)}$.
The quantum superplanes $\cA_{(M|N)}$ for $M\neq N$ are defined by 
setting the appropriate coordinates (and deformation parameters)
to zero in $\cA_{(\max(M,N)|\max(M,N))}$.
It is easily seen that the solutions \eqBMVcg\ and \eqBMVca,
obtained by solving the Yang-Baxter equations, are
fully compatible with the associativity constraints; i.e.\ 
imposing associativity does not lead to additional relations on the 
coordinates.  In general we expect this to be the case for all 
solutions to \eqBMVae\ 
satisfying the naturalness conditions \eqBMVd, \eqBMVk\ and \eqBMVl.

Now we briefly discuss the differential calculus on a fused quantum
superplane.  We start with the formulation in terms of $\wcR$. 
One can introduce an exterior derivative $d$, differentials $dz^A$
and derivatives $\p_A$, such that 
\eqn\eqBMVba{ 
d^2 ~=~ 0\,, \qquad d ~=~ dz^A\,\p_A\,,
}
and for forms $\om_1\in \mywedge^k \wcA_q ,\, \om_2\in \mywedge^l \wcA_q$
\eqn\eqBMVbb{
d (\om_1 \om_2) ~=~ (d\om_1) \om_2 + (-1)^k \om_1 (d\om_2)\,.
}
(We have chosen the convention where $d$ commutes with both the 
bosonic and fermionic coordinates of the quantum superplane.)
Following the steps in [\WZ] one finds that the following exchange 
formulas provide a consistent differential calculus on the fused 
quantum superplane $\wcA_q$
\eqn\eqBMVbc{ \eqalign{
dz^A z^B & ~=~ \wcR^{AB}_{CD} \, z^C dz^D \,, \cr
dz^A dz^B & ~=~ - \wcR^{AB}_{CD} \, dz^C dz^D \,, \cr
\p_C z^A & ~=~ \de^A_C + \wcR^{AB}_{CD}\, z^D \p_B\,, \cr
\p_C \p_D & ~=~ \wcR^{BA}_{CD} \, \p_A \p_B\,, \cr
dz^A \p_C & ~=~ \wcR^{AB}_{CD} \p_B dz^D \,.\cr}
}
To recover the differential calculus on $\cA_q$, we
have to specialize the formulas in \eqBMVbc\ to the hyperplane $z^0=1$.
Since $\wcR^{0B}_{CD} = \de^B_C \de^0_D$ and 
$\wcR^{A0}_{CD} = \de^A_D \de^0_C$,
it is easily seen that the exchange formulas are consistent with the 
assignment $z^0=1,\, dz^0=0$.  Also, when considering forms on the 
hyperplane $\cA_q$, there is no need to define the derivative $\p_0$.
However, for consistency, we need to show that $\p_0$ does not occur 
on the right hand side if it does not occur on the 
left hand side of \eqBMVbc.
Again, this follows from the equations $\wcR^{0B}_{CD} = \de^B_C \de^0_D$ and 
$\wcR^{A0}_{CD} = \de^A_D \de^0_C$.
(Sometimes, it can happen that $\p_0$ can be eliminated by combining 
various equations of \eqBMVbc\ in which $\p_0$ does occur.  In these 
cases, however, at least in examples, no independent equations on the
other derivatives are obtained.)
Thus, we conclude that \eqBMVbc\ leads to a consistent 
differential calculus on $\cA_q$.

A possible application of the foregoing might be a new kind of
superspace supersymmetry and supergravity, or two-dimensional
superconformal field theory, where the Grassmann thetas are replaced by
our kind of Clifford thetas.  One should then work out how the deformed 
supersymmetry algebra acts on the supercoordinates, using methods
which are quite common in quantum groups.
These ideas remain to be explored.

\bigskip\bigskip
\leftline{\bf Acknowledgements}

PvN would like to thank the Institute for Theoretical Physics of the
University of Adelaide for hospitality while part of this work was being done.
We would like to thank Jan de Boer for discussions.
PB and JM acknowledge the support of the Australian Research Council,
while PvN is supported by NSF grant PHY9309888.

\footatend\immediate\closeout\rfile
\baselineskip=14pt{\bigskip\noindent {\bf References}}%
\bigskip{\frenchspacing%
\parindent=20pt\escapechar=` \input refs.tmp\vfill\eject}\nonfrenchspacing
\vfil\eject\end